\documentclass[12pt]{article}

\usepackage[T1]{fontenc}

\usepackage{amsmath}
\usepackage{amssymb}
\usepackage{amsfonts}
\usepackage{bm}
\usepackage{graphicx}
\usepackage[dvipsnames]{xcolor}
\usepackage[colorlinks=true,urlcolor=blue,linkcolor=blue,citecolor=blue]{hyperref}
\usepackage{cite}
\usepackage{dsfont}
\usepackage{array}

\usepackage[top=1.135in,bottom=1.17in,left=1.16in,right=1.16in]{geometry}
\numberwithin{equation}{section}

\allowdisplaybreaks

\newcommand{\beq}{\begin{equation}}
\newcommand{\eeq}{\end{equation}}
\newcommand{\beqa}{\begin{eqnarray}}
\newcommand{\eeqa}{\end{eqnarray}}

\renewcommand{\d}{{\partial}}
\newcommand{\dd}{{\rm d}}

\newcommand{\BR}{{\rm BR}}

\begin{document}
\thispagestyle{empty}

\begin{center}

{\bf\Large   Cosmology of an Axion-Like Majoron}

\vspace{30pt}

A.J.~Cuesta,$^{(1)}$ M.E.~G\'omez,$^{(2)}$ J.I.~Illana,$^{(3)}$ M.~Masip$^{(3)}$ 

\vspace{15pt}

{\it $^{(1)}$Departamento de F{\'\i}sica} \\
{\it Universidad de C\'ordoba, E-14071 C\'ordoba, Spain}
\vspace{10pt}

{\it $^{(2)}$Departamento de Ciencias Integradas} \\
{\it Universidad de Huelva, E-21071 Huelva, Spain}
\vspace{10pt}

{\it $^{(3)}$CAFPE and Departamento de F{\'\i}sica Te\'orica y del Cosmos} \\
{\it Universidad de Granada, E-18071 Granada, Spain}

\vspace{15pt}

{\tt ajcuesta@uco.es, mario.gomez@dfa.uhu.es, jillana,masip@ugr.es}

\today

\vspace{25pt}

\end{center}

\begin{abstract}
We propose a singlet majoron model that defines an inverse seesaw mechanism in the $\nu$ sector. 
The majoron $\phi$ has a mass $m_\phi\approx 0.5$ eV and a coupling to the $\tau$ lepton  
similar to the one to neutrinos. In the early universe it is initially in thermal
equilibrium, then it decouples at $T\approx 500$ GeV and 
contributes with just $\Delta N_{\rm eff}=0.026$ during BBN. At $T=26$ keV (final stages of BBN) a primordial 
magnetic field induces resonant 
$\gamma \leftrightarrow \phi$ oscillations that transfer 6\% of the photon energy into majorons,
implying $\Delta N_{\rm eff}=0.55$ and a 4.7\% increase in the baryon to photon ratio.
At $T\approx m_\phi$ the majoron 
enters in thermal contact with the heaviest neutrino and it finally decays 
into $\nu \bar \nu$ pairs near recombination, setting  $\Delta N_{\rm eff}=0.85$. 
The boost in the expansion rate at later times 
may relax the Hubble tension (we obtain $H_0=(71.4\pm 0.5)$ km/s/Mpc), 
while the processes $ \nu\bar \nu \leftrightarrow \phi$ 
suppress the free streaming of these particles
and make the model consistent with large scale structure observations. Its lifetime and 
the fact that it decays into neutrinos instead of photons 
lets this axion-like majoron avoid the strong bounds that affect  other
axion-like particles of similar mass and coupling to photons.

\end{abstract}

\newpage

\section{Introduction\label{sec1}}

Neutrinos are a crucial ingredient of the $\Lambda$CDM cosmological model. Their mass, their interactions, or 
the presence of additional (sterile) modes may have implications on primordial nucleosynthesis (BBN), the 
cosmic microwave background 
(CMB) or the formation of large scale structures (LSSs) \cite{Lesgourgues:2018ncw}. 
In astrophysics they play an equally important role, 
for example, in the dynamics of stars like the Sun or in proto-neutron stars, where 
the neutrino sector is probed up to 
energies around 100 MeV \cite{Bahcall:1989ks}.
On the other hand, some fundamental questions about their nature 
remain unanswered. Are they Dirac particles that get massive at the electroweak (EW) scale, just like the rest of standard fermions, or is their mass revealing a new scale in particle physics? It is apparent that the search for answers 
can benefit from a large variety of different observations.

From a model-building point of view, the so called seesaw mechanism  \cite{Yanagida:1979as,GellMann:1980vs}
provides a very simple and minimal completion of the sector.
It explains that neutrinos are light because of a large hierarchy between the EW scale ($v$)
and the scale ($M_R$) where their right-handed partners, unprotected by chirality, get their mass. 
The
parameters (neutrino masses and mixings) that define this setup
are able to explain well the data \cite{deSalas:2020pgw}, 
despite some persistent anomalies in baseline \cite{Athanassopoulos:1996jb,AguilarArevalo:2007it}
and flavor \cite{Bigaran:2019bqv} experiments or 
possible difficulties in the modelling of supernova explosions \cite{Janka:2016fox,Rembiasz:2018lok}. 
One should keep in mind, however, that the neutrino sector
is basically {\it dark} and that the data leaves still plenty of room for departures from minimality. In this context, cosmology 
is becoming a powerful guide, signaling the new physics that could be favored: 
CMB \cite{Planck:2018nkj} or BBN \cite{Steigman:2007xt} observables have been measured at precision levels, 
whereas the volume of data on LSS 
\cite{eBOSS:2020yzd,DES:2021wwk} will increase substantially in upcoming experiments like 
EUCLID \cite{Laureijs:2011gra} or J-PAS \cite{Benitez:2014ibt}. 
Indeed, the Lithium problem (BBN implies 2--3 times more primordial 
Li than observed \cite{Cyburt:2015mya}) 
or the $H_0$ tension (type-Ia supernovae calibrated on Cepheids \cite{Riess:2020fzl}
and strong lensing \cite{Wong:2019kwg} suggest a 4--6 $\sigma$ larger expansion
rate than predicted by CMB and BAO data) 
could be related to the physics of neutrinos.

Here we will explore a different completion of the neutrino sector that does not imply minimality 
\cite{Chikashige:1980ui,Chacko:2003dt}. We propose a singlet majoron model where two different scales appear 
in a single phase transition: a scale $v_X\approx 1$ TeV that breaks a global $U(1)_X$ symmetry, plus
a much smaller scale $\epsilon \, v_X$ that breaks a $Z_3\subset U(1)_X $ subgroup of discrete symmetries. 
The value of $m_\nu$ is then explained by the 
spontaneous breaking of the discrete symmetry, $m_\nu \propto \epsilon \, v_X$, like in inverse seesaw models 
\cite{Mohapatra:1986bd,Bernabeu:1987gr,Bolton:2019pcu,Hernandez-Tome:2019lkb,Hernandez-Tome:2020lmh}, 
whereas all the extra particles except for the majoron $\phi$ get masses of order $v_X$.

Our scenario is a variation of the one proposed in \cite{Escudero:2019gvw,Escudero:2021rfi} 
(see also \cite{ift}); in particular, it implies the same type of majoron coupling  to neutrinos, 
$\lambda_\nu\approx m_\nu/v_X$. However, our proposal can also accommodate tiny couplings to the 
charged leptons that induce the dimension-5
operator $\phi \widetilde F_{\mu \nu} F^{\mu \nu}$: this is an axion-like majoron (ALM). 

\section{The model \label{sec2}}

Consider an extension of the SM that includes three fermion singlets\footnote{We use a 2-component spinor notation with
all the fields of left-handed chirality; at the end of the section we will translate it to the usual 4-spinor notation.}
$\{ N,\,N^c,\,n \}$ plus several complex scalar singlets $\{ s_1,\,s_2,\,\hdots \}$. Let us assume that the model
is valid up to a cutoff scale
$\Lambda\approx 10$ TeV and that, up to gravitational effects \cite{Barbieri:1979hc}, 
it is invariant under the global $U(1)_X$ symmetry
\beq
\psi \to e^{-i\, Q_X\,\theta} \;\psi\,,
\eeq
with the charges $Q_X$ given in Table \ref{charges}. 
\begin{table}
\begin{center}
\begin{tabular}{|c|ccccccc|}
\hline
$U(1)_X$
 & \hspace{0.1cm} $(\nu\; e)$ \hspace{0.1cm}& \hspace{0.1cm} $e^c$  \hspace{0.1cm} & 
 \hspace{0.1cm} $N$ \hspace{0.1cm}  & 
 \hspace{0.1cm} $N^c$  \hspace{0.1cm} & \hspace{0.1cm} $n$  \hspace{0.1cm} 
 & $(h^+\; h^0)$  & \hspace{0.1cm} $s_1,\, s_2,\,\hdots$  \hspace{0.1cm} \\ [0.4ex]
\hline 
$Q_X$ & $+1$ &
$-1$ & $-2$ & $-1$ & $0$ & $0$ & $1,\,2,\,\hdots$ \\ [0.4ex]
\hline
\end{tabular} 
\caption{Charges under the global symmetry.}
\label{charges}
\end{center}
\end{table}
Then a linear 
combination $s$ of the singlets $s_i$ gets a VEV $v_X$ and breaks the global symmetry:
\beq
s={1\over \sqrt{2}} \left( v_X + r \right) e^{i\,\phi/v_X}\,,
\eeq 
with $\phi$ the Goldstone boson. We will suppose that the scalar potential may define $s$ as {\it any} combination
of singlet flavors, including combinations with a tiny component along a given flavor. 
For example, for two scalars $s_1$ and $s_2$ the potential
\beq
V=-m_1^2\; s_1^\dagger s_1 + \alpha \left(  s_1^\dagger s_1 \right)^2 + \left|  \beta\, s_1^2 - m_2\, s_2 \right|^2
\eeq
is invariant under $U(1)_X$. If we take $\beta \ll \alpha$ and $m_1\approx m_2$ this potential implies a VEV for the flavor combination
$s\approx s_1 + \epsilon s_2$, with $\epsilon = \beta m_1/( \sqrt{2\alpha}\, m_2) \ll 1$. This means 
VEVs $\langle s_1\rangle = O(m)$ and $\langle s_2\rangle = O(\epsilon \,m)$, a majoron $\phi$ with components 
along both flavors, $\phi\approx \phi_1 + \epsilon \phi_2$, and masses of $O(m)$ for 
the rest of degrees of freedom in $s_1$ and $s_2$. More complicated effective potentials 
could break the global symmetry along any combination
of singlet flavors and leave the majoron as the only scalar with a mass much smaller than $v_X$
($m_\phi$ will not be zero if the global symmetry is
not exact).

Let us then take the main component in $s$ along $s_3$, with charge $Q_X=3$. 
The terms relevant for neutrino masses and majoron couplings allowed by the symmetry are 
\beq
-{\cal L} \supset y_i \; h^0 \nu_i N^c + y'\; s_3 N N^c + {\Lambda_n\over 2} \; n n + {\rm h.c.}\,,
\label{yukawas}
\eeq
whereas the Yukawas 
$h^0 \nu N$, $h^0 \nu n$, $s_3 N N$, $s_3 N^cN^c$, $s_3 n N$, $s_3 n N^c$ and $s_3 n n$ are all 
forbidden.  The VEV $\langle s_3 \rangle$ will obviously break $U(1)_X$,
but we notice that a global transformation of parameter $\theta=-2\pi/3$ leaves the vacuum invariant:
\beq
\langle s_3 \rangle \to e^{- 3i\,(-2\pi/3)}\, \langle s_3 \rangle= \langle s_3 \rangle\,.
\eeq
The spinors in Table \ref{charges} change non-trivially under this transformation:
\beq
\nu\to \alpha \, \nu \hspace{0.7cm} N\to \alpha \, N \hspace{0.7cm}
N^c\to \alpha^* \, N^c \hspace{0.7cm} n\to +1\,n \,,
\eeq
with $\alpha = e^{i\,2\pi/3}$. Therefore, the effective Lagrangian must respect  to all order
this unbroken $Z_3$ symmetry. In particular, Higgs ($v/\sqrt{2}$) and singlet ($v_X/\sqrt{2}$) 
VEVs will not generate bilinears of type $\nu N$, $\nu n$, $N N$, $N^cN^c$, $n N$ or $n N^c$ through operators 
of any dimension (again, up to gravitational effects), as these terms would break the discrete symmetry.
The neutrino mass matrix reads then
\beq
-{\cal L} \supset {1\over 2}
\begin{pmatrix} \nu_1 & \nu_2 & \nu_3 & N & N^c & n \end{pmatrix}
\begin{pmatrix}
        \cdot   &  \cdot   &  \cdot   &  \cdot & m_1 &  \cdot \\
        \cdot   &  \cdot   &  \cdot   &  \cdot & m_2 &  \cdot \\
        \cdot   &  \cdot   &  \cdot   &  \cdot & m_3 &  \cdot \\
        \cdot   &  \cdot   &  \cdot   &  \cdot & M   &  \cdot \\
        m_1 & m_2 & m_3 & M &  \cdot  &  \cdot \\
        \cdot   &  \cdot   &  \cdot   &  \cdot &  \cdot  & \Lambda_n\\
\end{pmatrix}
\begin{pmatrix} \nu_1 \\ \nu_2 \\ \nu_3 \\ N \\ N^c \\ n \end{pmatrix} + {\rm h.c.},
\label{massmatrix1}
\eeq
where $m_i\equiv y_i v / \sqrt{2}$ and $M\equiv y' v_X / \sqrt{2}$. This is a rank 3 matrix that implies 
3 massless neutrinos, two neutrinos defining a 
Dirac field of mass $M'=\sqrt{M^2+m_1^2 + m_2^2 +m_3^2}$, heavy--light mixings $\theta_{N\nu}\approx m_i/M$, 
and a Majorana neutrino of mass $\Lambda_n$. The discrete symmetry protects the light neutrinos from a mass
$m_\nu\propto v^2/v_X$, which is the ratio of scales setting $m_\nu$ in seesaw scenarios. 
In addition, since under the $Z_3$ symmetry
$(\nu_i, N) \to \alpha \,(\nu_i, N)$ and $\phi_3 \to \phi_3$, 
a coupling $i\lambda_\nu\,\phi \nu \nu$ is
also forbidden.

To generate $\nu$ masses 
we need that the singlet VEV includes a small component breaking that symmetry.
A minimal possibility is 
\beq
s=s_3+\epsilon \, s_4 \; \implies \;  \langle s_4 \rangle = \epsilon\, \langle s_3 \rangle \;;\;\;\; \phi=\phi_3+\epsilon \, \phi_4\,,
\eeq
with $\epsilon \ll 1$. The VEV  $\langle s_4\rangle$ may then induce entries
everywhere in the matrix ${\cal M}$ in 
(\ref{massmatrix1}). In particular, we  obtain masses for the three light neutrinos using
one dimension 4 and two dimension 6 operators:
\beq
-{\cal L} \supset y_{IS}\,s_4 N N + 
\tilde y_i\, {s_3^\dagger \, s_4  \over \Lambda^2}
\begin{pmatrix} h^+ \\ h^0 \end{pmatrix}
\begin{pmatrix} \nu_i \\ e_i \end{pmatrix} N +
\tilde y_i'\, {s_3 \, s_4^\dagger  \over \Lambda^2}
\begin{pmatrix} h^+ \\ h^0 \end{pmatrix}
\begin{pmatrix} \nu_i \\ e_i \end{pmatrix} n \,.
\label{lagmass}
\eeq
The first term implies an entry $\mu=y_{IS} \sqrt{2} \,\epsilon v_X$ in ${\cal M}_{4\,4}$ that gives mass to one of the
neutrinos ($\nu'_3$; we use a prime to
indicate mass eigenstates) through an inverse seesaw mechanism,
\beq
m_{\nu'_3} \approx \mu \left( {m \over M} \right)^2
\eeq
with $m=\sqrt{m_1^2 + m_2^2 + m_3^2}$. In addition, $\nu'_3$
has now components along $N$ and also $N^c$,
\beq
\theta_{N\nu}\approx {m\over M} \;;\hspace{0.5cm} \theta_{N^c\nu}\approx {\mu\,m\over M^2}\;,
\eeq
that induce couplings of the majoron to $\nu'_3$ via two different operators:
\beq
y_{IS}\,s_4 N N \to i\, {m_{\nu'_3}\over 2 v_X}\, \phi\,\nu'_3 \nu'_3\;,\hspace{0.5cm}
y'\; s_3 N N^c \to i\, {m_{\nu'_3}\over v_X}\, \phi\,\nu'_3 \nu'_3\;.
\label{nuphi}
\eeq
Including the mass contributions from all the operators in (\ref{lagmass}) we obtain
\beq
-{\cal L} \supset {1\over 2}
\begin{pmatrix} \nu_1 & \nu_2 & \nu_3 & N & N^c & n \end{pmatrix}
\begin{pmatrix}
        \cdot   &  \cdot   &  \cdot   &  0 & 0 &  \tilde \mu' \\
        \cdot   &  \cdot   &  \cdot   &  \tilde \mu & 0 &  \cdot \\
        \cdot   &  \cdot   &  \cdot   &  \cdot & m &  \cdot \\
        0   &  \tilde \mu   &  \cdot   &  \mu & M   &  \cdot \\
        0 & 0 & m & M &  \cdot  &  \cdot \\
        \tilde \mu' & \cdot & \cdot &  \cdot &  \cdot  & \Lambda_n\\
\end{pmatrix}
\begin{pmatrix} \nu_1 \\ \nu_2 \\ \nu_3 \\ N \\ N^c \\ n \end{pmatrix} ,
\label{massmatrix}
\eeq
where we have redefined the three
active neutrinos $\nu_i$ so that $m_{1,2}=0$ and $\tilde \mu_1=0$.
The second light neutrino ($\nu'_2$) 
gets a mass $m_{\nu'_2}\approx \tilde \mu^2 /\mu$, while the mass $m_{\nu'_1}\approx \tilde \mu'^2/\Lambda_n$
of the lightest one is generated through a type I seesaw mechanism. 
It is easy to deduce that the couplings of $\nu'_{1,2}$ to the majoron are 
$\lambda_{\nu'_1}={m_{\nu'_1}/(2v_X)}$ and $\lambda_{\nu'_2}={m_{\nu'_2}/v_X}$. The rest of entries (dots in the
matrix above) are also 
suppressed by $\epsilon$ and inverse powers of the cutoff; they 
are not necessary to define the model, although they may introduce $O(1)$ corrections
to the $\nu$ masses and mixings.

Another aspect of our setup concerns the possible coupling of the majoron 
to charged leptons. Let us consider
$\ell=\mu,\,\tau$ (the coupling to electrons is severely constrained \cite{Raffelt:1994ry}); 
we assign $Q_X(\ell)=+1$ and $Q_X(\ell^c)=-1$ so that the usual Yukawa term
$\, h^0 \ell \ell^c\, $ 
is allowed by the symmetry. In turn, 
this implies that higher dimensional operators of type $\, s_3^\dagger s_3 h^0 \ell \ell^c\,$ or 
$\, s_4^\dagger s_4 h^0 \ell \ell^c\,$ will not introduce couplings 
to the majoron, just to the (massive) radial component in $s$. Unlike neutrinos, chiral 
charged leptons do not couple to the majoron.
We can, however, add a pair of heavy (vectorlike) lepton singlets
$(E_{1}, E^c_{-1})$ and $(E_{4}, E^c_{-4})$ with charges $Q_X=\pm 1$ and $Q_X=\pm 4$, 
respectively, and masses
$m_{E_{1,4}} > v_X$. These leptons may couple to the majoron through Yukawas like 
$y_E\, s_3 E_{1} E^c_{-4}$. In that case $\ell$ could also couple to the majoron via mixing 
with the heavy sector, {\it e.g.}, $s_3 h^0 \ell E^c_{-4}$ induces a bilinear $m'\,\ell E^c_{-4}$ and 
a term 
\beq
-{\cal L} \supset i \lambda_\ell \; \phi \,\ell' \, {\ell^c}' + {\rm h.c.}
\label{ephi}
\eeq
of order $\lambda_\ell \approx m'/ v_X\, (m_\ell /m_E)^2$.

It may be useful to translate this 2-spinor notation to the usual one with 4-spinors 
in the chiral representation. All the neutrinos 
discussed here can be arranged as the left-handed component of a self-conjugate (Majorana) field,
\beq
\nu_{i=1,2,3} \equiv \begin{pmatrix} {\nu_i} \\ {\bar \nu_i} \end{pmatrix}\;;\;\;\;
\nu_4 \equiv \begin{pmatrix} N \\ \bar N \end{pmatrix}\;;\;\;\;
\nu_5 \equiv \begin{pmatrix} N^c \\ \bar N^{c} \end{pmatrix}\;;\;\;\;
\nu_6 \equiv \begin{pmatrix} n \\ \bar n \end{pmatrix}\;,
\eeq
where bi-spinor indexes ($\psi_\alpha$ or $\bar \psi^{\dot{\alpha}}$) are omitted. 
For the charged leptons we have
\beq
\ell \equiv \begin{pmatrix} \ell \\ \bar \ell^{c} \end{pmatrix}\;.
\eeq
Dropping the prime, the majoron couplings 
in (\ref{nuphi}) and (\ref{ephi})
would read
\beq
{\cal L} \supset i \,\lambda_{\nu_i} \, \phi \; \overline {\nu_i} \, \gamma_5  \,\nu_i + 
i\, \lambda_\ell \, \phi \; \overline {\ell} \, \gamma_5  \,\ell \,.
\label{lnu}
\eeq
Actually, in our model the neutrino basis with diagonal majoron couplings used above
does not need to coincide with the basis of mass eigenstates, as some of the 
operators contributing to the mass have no axial dependence. For simplicity, however, we will take diagonal couplings and $\lambda_{\nu_i}\approx m_{\nu_i}/v_X$.

\begin{figure}
\begin{center}
\raisebox{-0.5\height}{\includegraphics[scale=0.62]{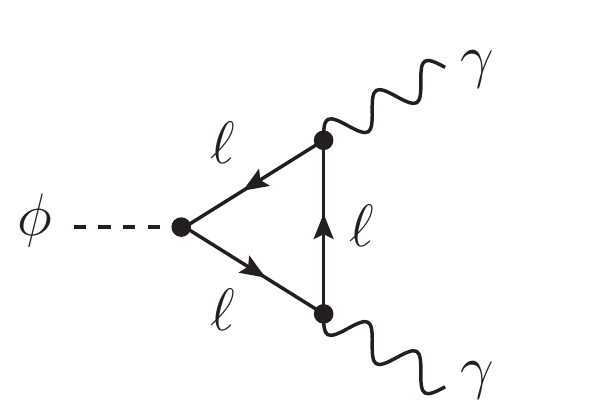}} \hspace{0.1cm}
\raisebox{-0.5\height}{\includegraphics[scale=0.62]{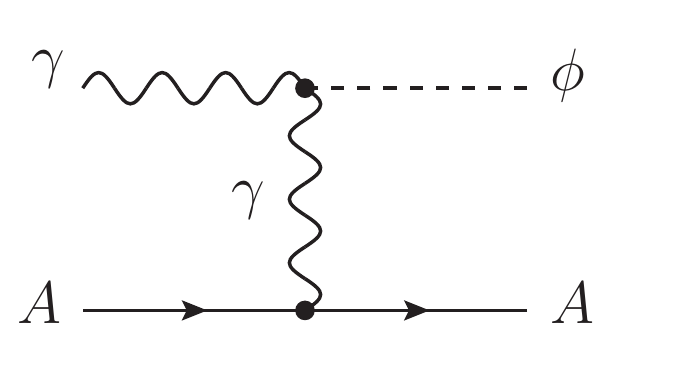}}
\raisebox{-0.5\height}{\includegraphics[scale=0.62]{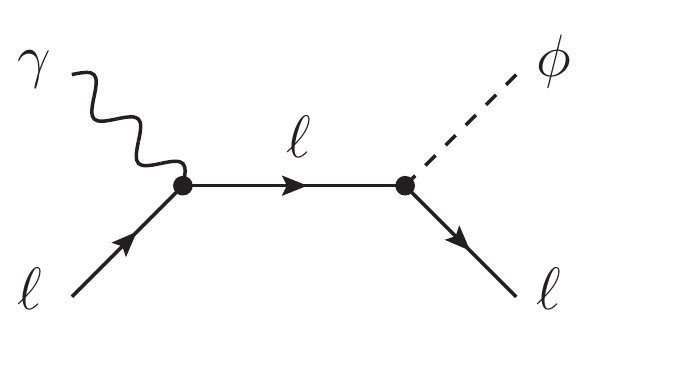}}
\caption{One-loop contribution to $g_{\phi\gamma\gamma}$ (a); ALM production through Primakoff (b) and Compton (c) processes (a $u$-channel diagram in Compton is omitted).}
\label{fig1}
\end{center}
\end{figure}

A final observation concerns the axion-like character of the majoron in this model.
The coupling $\lambda_\ell$ with a charged lepton discussed above induces at one loop (see Fig.~\ref{fig1}a) the operator
\beq
{\cal L} \supset -{1\over 4} \,g_{\phi\gamma\gamma}\, \phi \widetilde F_{\mu\nu} F^{\mu\nu}\,,
\eeq
with
\beq
g_{\phi\gamma\gamma}=-{\alpha \lambda_\ell \over 4\pi \tau m_\ell} 
\left( {\rm Li}_2 \! \left( {2\over 1 + \sqrt{1-\tau^{-1}}} \right) +
{\rm Li}_2\!  \left( {2\over 1 - \sqrt{1-\tau^{-1}}} \right) \right) \approx - {\alpha \lambda_\ell \over 2\pi m_\ell} \,,
\label{loop}
\eeq
an approximation that is valid when $\tau = m_\phi^2/(4m_\ell^2) \ll 1$ ({\it i.e.}, once we integrate out the charged lepton). 
Notice that there may be other contributions to $g_{\phi\gamma\gamma}$ from, for example, the heavy charged 
lepton singlets introduced
above. Therefore, in this model a given value of $g_{\phi\gamma\gamma}$ just implies  
$\lambda_\ell \le - (2\pi m_\ell g_{\phi\gamma\gamma})/\alpha$, where the equality holds  
when the dominant contribution 
comes from the diagram in Fig.~\ref{fig1}.
In any case, the axion-like coupling $g_{\phi\gamma\gamma}$
will be determinant in the cosmological evolution of the ALM.

\section{Cosmological evolution: $T>10$ eV}

Let us summarize the model under study and choose definite values for all its parameters. 
As discussed in the previous section, the model implies 
the presence of three heavy neutrinos; we will take a quasi-Dirac pair with  $M_{4,5}\approx 500$ GeV 
and $M_6\approx 1$ TeV for the Majorana field. These neutrinos have unsuppressed 
Yukawa interactions [see Eq.~(\ref{yukawas})], so they will be in thermal equilibrium with the 
cosmic plasma at 
$T\ge 500$ GeV. For the majoron, also in equilibrium at these temperatures, we take
\beq
m_\phi= 0.5\;{\rm eV}\;;\hspace{0.5cm} 
g_{\phi\gamma\gamma}=1.46\times 10^{-11}\;{\rm GeV}^{-1}
\eeq
and 
\beq
\lambda_{\nu_3}=6.8\times 10^{-14}\;;\hspace{0.5cm} \lambda_{\nu_2}=1.3\times 10^{-14}\;;\hspace{0.5cm} 
\lambda_{\nu_1}=0.49\times 10^{-14}\,.
\label{lambdas}
\eeq
The chosen value of $g_{\phi\gamma\gamma}$ avoids astrophysical bounds 
\cite{Masso:1995tw,Raffelt:2006cw,Ayala:2014pea} and fixes $\lambda_\mu\le 1.3\times 10^{-9}$ 
and $\lambda_\tau\le 2.2\times 10^{-8}$. 
Because of its tiny mass, the majoron cannot decay into these charged leptons, just into photon or neutrino pairs:
\beq
\Gamma (\phi \to \nu_i \bar \nu_i)={\lambda_{\nu_i}^2\over 4\pi}\,m_\phi\,\sqrt{1-{4 m_{\nu_i}^2\over m_\phi^2}}\;;\hspace{0.5cm}
\Gamma (\phi \to \gamma\gamma)= {g_{\phi\gamma\gamma}^2\over 64\pi}\,m_\phi^3\,.
\eeq
However, for the couplings that we have assumed the branching ratio into gammas is negligible (around $10^{-15}$): 
the majoron decays always into neutrinos with a lifetime
\beq
\tau_\phi = 3.5\times 10^{12} \; {\rm s} \,.
\eeq
The branching ratios BR$_{\nu_i}$ of $\phi\to \nu_i\bar \nu_i$ to the three flavors are
\beq
{\rm BR}_{\nu_3}=0.96\,,\;\; 
{\rm BR}_{\nu_2}=0.035\,,\;\; 
{\rm BR}_{\nu_1}=0.005\,.
\label{BRs}
\eeq

To reconstruct the cosmological history of the ALM we need the Hubble parameter $H$ 
at each temperature:
\beq
H^2 = {8\pi \over 3 M_P^2}\,\rho \, .
\label{H2}
\eeq
Let us focus on  
$T>10$ eV, when the universe is still radiation-dominated. The energy density can be expressed in terms of 
the total number of effectively massless degrees of freedom ($g_*$):
\beq
\rho_R={\pi^2\over 30}\,g_* \, T^4\,.
\eeq
We will take the standard values of $g_*$ in \cite{Husdal:2016haj}.
At $T\approx 500$ GeV all the SM degrees of freedom plus the heavy neutrinos and the majoron are in thermal equilibrium and
contribute to $g_*$. Then the temperature drops below $M_{4,5}$ and the heavy neutrinos disappear, 
transferring their entropy to the thermal bath. Our initial universe consists of a plasma with all the standard
particles ($g_*=106.75$) and the ALM ($\Delta g_*= 1$), all of them with the same temperature.

Once the heavy neutrinos have disappeared the majoron goes out of equilibrium and its abundance freezes out. 
Primakoff collisions $\gamma A \to \phi A$ with
$A$ any charged particle in the plasma (see 
Fig.~\ref{fig1}b) were in equilibrium at $T>1$ PeV \cite{Masso:1995tw} but 
are negligible at $T< 500$ GeV, whereas 
Compton-like processes (in Fig.~\ref{fig1}c) are also ineffective \cite{Cadamuro:2010cz}. In particular, we estimate 
$\Delta \rho_\phi \le 5\times10^{-4} \rho_\phi^{\rm eq}$ from Compton 
collisions at $T\approx m_\mu$ if the dominant contribution in Eq.~(\ref{loop})
came from the muon loop or $\Delta \rho_\phi \le 8\times 10^{-3} \rho_\phi^{\rm eq}$ if it came from
$\ell=\tau$. All the
interactions of the majoron with the neutrinos are negligible as well. Therefore,
as $T$ decreases the entropy in the heavy
degrees of freedom of the SM goes into the lighter ones, but not into majorons. This produces
a difference in the temperature in both components of the plasma. At $T\approx 1$ MeV, right
before $\nu$ decoupling, $e^+ e^-$ annihilation and BBN, we have
\beq
T_\phi = \left( {g_{*s}(1\;{\rm MeV}) \over g_{*s}(200\;{\rm GeV})} \right)^{\!1/3} T = 0.463\, T\,,
\eeq
where $T$ is still the temperature of both photons and neutrinos and $g_{*s}\approx g_{*}$ \cite{Husdal:2016haj}
does not include the majoron. 
We may give the contribution of the majoron
to $\rho_R$ in terms of $N_{\rm eff}$,
\beq
\rho_\nu + \rho_\phi = {7 \pi^2\over 120} \, N_{\rm eff} \, T_\nu^4\,,
\eeq
with 
\beq
\Delta N_{\rm eff}= {4\over 7} \left( T_\phi \over T_\nu \right)^4 = 0.026\,.
\eeq

At $T<1$ MeV the majoron-neutrino sector is decoupled from photons and electrons, so
this $\Delta N_{\rm eff}$ already present at the beginning of BBN (plus a later 0.044 contribution from 
$e^+ e^-$ annihilation and other subleading effects, see \cite{Escudero:2020dfa,Bennett:2020zkv}) 
could evolve unchanged down to low temperatures.
The ALM model, however, admits a very different possibility: the resonant conversion 
of a fraction of CMB photons into majorons \cite{Raffelt:1987im,Mirizzi:2009nq} 
at the end of BBN. 
This possibility rests on two basic observations. First, photons in a medium get a mass that can be  
expressed in terms of the index of refraction
describing their propagation. In the cosmic plasma the main contribution to this mass 
comes from their interaction with free (not bounded in atoms) electrons. Second, the presence of a background 
magnetic field mixes the photon with the axion-like particle; this separates mass and interaction eigenstates and may
produce oscillations that become resonant when the two masses coincide.

Let us then assume the presence of a primordial magnetic field $\bm{B}$ with cosmological coherence length ($\lambda_0\ge 1$ Mpc)
and field lines frozen in the plasma, so that  $B\approx B_0 \,(T/T_0)^2$. 
We will take\footnote{The effect to be discussed is proportional to $g_{\phi \gamma\gamma} B_0\approx 4\times 10^{-11}$ GeV$^{-1}$nG.}
 $B_0=3$ nG \cite{Planck:2015zrl}
and assume that the strength of the magnetic field is similar everywhere, although its direction may change in different domains.
We will follow the density matrix formalism developed by Ejlli and Dolgov (ED) in \cite{Ejlli:2013uda}, that 
accounts for the breaking of coherence (photon interactions that interrupt the
oscillations) and provides a very simple estimate when the dominant contribution is at the resonant temperature.

At $T<m_e$ the effective operators involved in the photon-ALM oscillations are
\beqa
{\cal L}\supset \!\! &-& \!\!{1\over 4} F_{\mu\nu} F^{\mu\nu}-{g_{\phi\gamma\gamma}\over 4} \phi\widetilde F_{\mu\nu} F^{\mu\nu}
+{1\over 2} \left( \d_\mu \phi \d^\mu \phi -m_\phi^2 \phi^2 \right)
\nonumber \\
\!\! &+& \!\!{\alpha^2\over 90m_e^4}\left( ( F_{\mu\nu} F^{\mu\nu} )^2 + {7\over 4} 
( \widetilde F_{\mu\nu} F^{\mu\nu} )^2 \right)\,.
\eeqa
Notice that the external field $\bm{B}$ will also give an effective mass $m_{\rm QED}$ to the photon
through the dimension-8 Euler-Heisenberg operator. In the WKB approximation the 
coupled equations of motion reduce to \cite{Raffelt:1987im}
\beq
\left( \left(\omega + i\d_{\bm{x}} \right) {\mathds{1}} + 
\begin{pmatrix} m_+ & 0 & 0 \\ 0 & m_\times & m_{\phi \gamma} \\ 0 & m_{\phi \gamma} & m_a \end{pmatrix}\right)
\begin{pmatrix} A_+ \\ A_\times \\ \phi \end{pmatrix} = 0\,,
\eeq
where $\omega$ is the photon energy, $\bm{x}$ its direction of propagation, $A_+$ and $A_\times$ correspond, respectively, to
the polarizations perpendicular and parallel to $\bm{B}$, $m_{+,\times}= \omega(n-1)_{+,\times}$ with $n$ the total refractive index, 
$m_{\phi \gamma}= g_{\phi\gamma\gamma} B_T/2$ with $B_T$ the component of $\bm{B}$
orthogonal to $\bm{x}$, and $m_a=-m_\phi^2/(2\omega)$. At a temperature $T$, for a photon of energy $xT$ these 
mass parameters are \cite{Ejlli:2013uda}
\beqa
m_{\phi\gamma}(T) \!\!&=&\!\! 1.10\times 10^{-13} \left( {T\over {\rm GeV}} \right)^2\; {\rm GeV}\,;\nonumber\\
m_{\times}(T) \!\!&=&\!\! \left( 193 \,x \left( {T\over {\rm GeV}} \right)^5 
-75.9 \,x^{-1} \left( {n_e(T)\over T \;{\rm GeV}^2}\right) \right)\; {\rm GeV}\,;\nonumber\\
m_{a}(T) \!\!&=&\!\! - \,1.07\times 10^{-19} \,x^{-1} \left( { {\rm GeV} \over T} \right)\; {\rm GeV}\,,
\eeqa
where $m_{\times}=(m_{\rm QED} + m_{\rm pla})_\times$ and 
\beq
n_e(T)\approx 4\left({m_e T\over 2\pi} \right)^{\!3/2}\exp\left( -{m_e\over T} \right)+ 0.88\,\eta_B \,n_\gamma(T)\,,
\eeq
with $\eta_B=6\times 10^{-10}$ and $n_\gamma=\left( 2\zeta(3)/\pi^2\right)T^3$. ED obtain an analytic solution 
for the transition probability when the universe crosses the resonant temperature $\overline T$ where 
$\Delta m\equiv m_\times - m_a = 0$ (see Fig.~\ref{fig2}). The approximation is valid 
if {\it (i)} $\rho_\phi \ll \rho_\gamma\approx \rho_\gamma^{\rm eq}$  and 
{\it (ii)} the ALM interaction rate is $\Gamma_\phi\approx 0$. They express the result in terms of 
the occupation number $n_\phi(x,T)$ of the energy level $xT$ relative to the occupation number of photons in equilibrium:
\begin{figure}
\begin{center}
\raisebox{-0.5\height}{\includegraphics[scale=0.62]{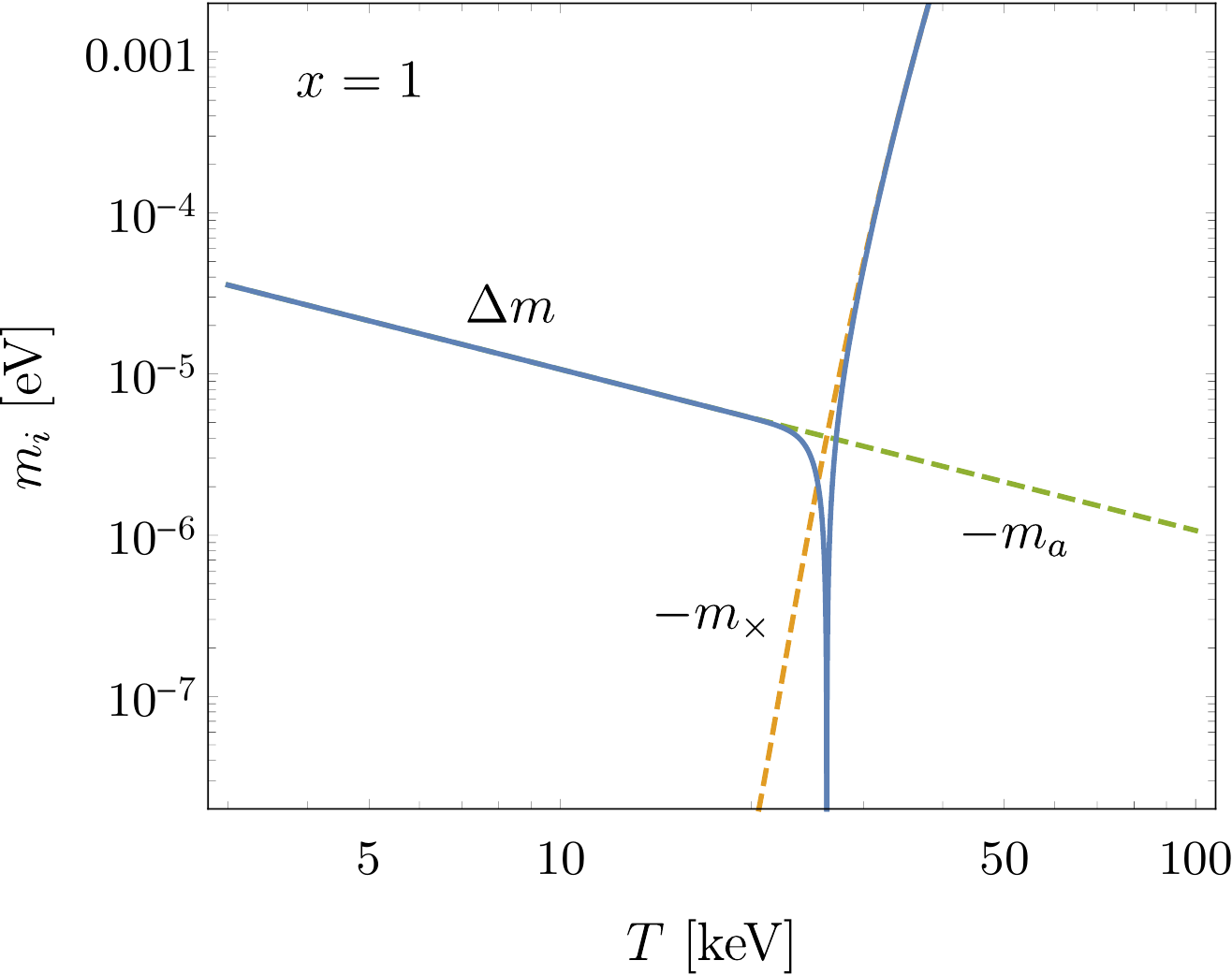}}
\caption{Masses for a photon/axion of $\omega= xT$ at temperatures near the resonance.}
\label{fig2}
\end{center}
\end{figure}
\beq
P_{\phi}(x,T)\equiv {\Delta n_\phi \over n_\gamma^{\rm eq}}\approx - 
\left( {2\pi\over 3 H } \right) \left. {m_{\phi \gamma}^2\over m_a + m_{\rm QED}}\right|_{T=\overline T}\,.
\eeq
For our choice of 
parameters we have $m_{\rm QED}\ll |m_{\rm pla}|$ and then a $\overline T$ that is independent from $x$, {\it i.e.}, 
all photons oscillate at the same temperature:
\beq
\overline T= 26\; {\rm keV}\,.
\eeq
The probability of oscillation, however, is proportional to the photon energy.
We obtain that 4.4\% of the photons, carrying a 6.3\% of the photon energy, convert into majorons
at $T\approx \overline T$. After this resonant conversion the effective mass of the photon becomes fast much smaller 
than $m_{\phi}$ and the number of ALMs freezes again. Notice that afterwards all photons rethermalize 
and the spectral distortion is completely erased.

In the simplified scenario just outlined the net effect is a sudden 1.6\% drop in the
photon temperature
and a 4.7\% increase in the baryon to photon ratio respect to the BBN value. 
Since neutrinos are not affected, at $T<\overline T$ the ratio $T_\gamma/T_\nu$ becomes a 1.6\% smaller than in the SM:
\beq
\left. {T_\gamma\over T_\nu} \right|_{SM}= \left({11\over 4}\right)^{1/3} 
\implies
{T_\gamma\over T_\nu}=
\left( 1-r_\gamma \right)^{1/4} \left({11\over 4}\right)^{1/3} \,.
\eeq
where $r_\gamma=0.063$ is the fraction of photon energy transferred to majorons.
In addition to the effective 
$3.044$ neutrino species, now the cosmic radiation also includes these majorons,
\beq
\rho_\phi^{(1)}= r_\gamma \,{\pi^2\over 15} \, {T_\gamma^4\over 1-r_\gamma}\,;\;\;\;
n_\phi^{(1)}=0.041\,{2\,\zeta(3)\over \pi^2} \, {T_\gamma^3\over (1-r_\gamma)^{3/4}}\,,
\eeq
plus the initial thermal population of majorons at  
\beq
T_\phi^{(2)}={0.463\over \left({11\over 4}\right)^{1/3} (1-r_\gamma)^{1/4}} 
\implies
\rho_\phi^{(2)}= {\pi^2\over 30}\, T_\phi^{(2)\,4}\,.
\eeq
We may express $\rho_R=\rho_\gamma+\rho_\nu+\rho_\phi$ at $T_\gamma < 26$ keV as
\beq
\rho_R = {\pi^2\over 15}\, T_\gamma^4 + {7\over 8}\,{\pi^2\over 15} \left( T_\gamma\over \left({11\over 4}\right)^{\!1/3}
\right)^{\! 4} \left( 3 + \Delta N_{\rm eff} \right),
\eeq
with 
\beq
\Delta N_{\rm eff} = 0.577\,.
\eeq
This radiation will evolve down to $T_\gamma< 10$ eV until 
neutrinos and majorons enter in 
thermal contact  at $T\approx m_\phi$
through decays and inverse decay collisions \cite{Chacko:2003dt,Escudero:2019gvw,Escudero:2021rfi}:
\beq
\nu \bar \nu \leftrightarrow \phi\,.
\eeq

\section{Cosmological evolution: $T<10$ eV} 

Let us analyze the final stage of the ALM evolution: from 10 eV to past recombination at  $T_\gamma \approx 0.26$ eV,
when all majorons have already decayed. 

As described in the previous section, the population of majorons that results from  photon oscillations
does not have a thermal distribution. We may, however,
find a temperature $T_\phi$ and a chemical potential $\mu_\phi$ defining a distribution that
reproduces the actual values of both
$\rho_\phi$ and $n_\phi$. In particular, we evolve $\rho_\phi=\rho_\phi^{(1)}+ \rho_\phi^{(2)}$ 
and $n_\phi=n_\phi^{(1)}+ n_\phi^{(2)}$ down to $T_\gamma=10$ eV and obtain  
\beq
T_\phi= 0.72\,T_\gamma\;;\;\;\; 
\mu_\phi= -0.41\,T_\gamma\;.
\eeq
In a similar way, we evolve the number and energy density of 
the three neutrino species; including the small
(non-thermal) component from $e^+ e^-$ annihilations we get
\beq
T_\nu= 0.72\,T_\gamma\;;\;\;\; 
\mu_\nu= 0.13\,T_\gamma\;.
\eeq
For the initial population of neutrinos we will take the same proportion of the three mass 
eigenstates\footnote{This is an example of scenario where the Quantum Mechanics approach fails
\cite{Falkowski:2019kfn} and we must work with neutrino mass eigenstates.}. 

We will follow the procedure described by Escudero in  \cite{Escudero:2020dfa}\footnote{The code 
NUDEC\_BSM \cite{Escudero:2020dfa,Escudero:2018mvt} is publicly available.} adapted to a majoron 
with stronger coupling to heavier neutrinos. The thermodynamic evolution of
a $\phi$--$\nu$ fluid in thermal contact through $\nu \bar \nu \leftrightarrow \phi$  can be expressed in terms of the 
decay width
\beq
\Gamma_\phi\approx {\sum_i \lambda^2_{\nu_i}\over 4\pi}\,m_\phi=3.7\times 10^{-28}\,m_\phi
\eeq
and the three branching ratios in Eq.~(\ref{BRs})\footnote{Notice that our definition of $\lambda_\nu$ in (\ref{lnu}) 
differs by a factor of $2$ from the one in
\cite{Escudero:2020dfa} and that $\Gamma_\phi$ refers there to the partial decay width.}. 
In particular, for the majoron we have
\beq
\d_t n_\phi = {\Gamma_\phi m_\phi^2 \over 2 \pi^2} \left[ \sum_i \BR_{\nu_i} \,T_{\nu_i} \,e^{2{\mu_{\nu_i}\over T_{\nu_i}}}\,
K_1\Big({m_\phi\over T_{\nu_i}}\Big) -  T_\phi \,e^{\mu_\phi\over T_\phi}\,K_1\Big({m_\phi\over T_\phi}\Big) \right]\,;
\eeq
\beq
\d_t \rho_\phi = {\Gamma_\phi m_\phi^3 \over 2 \pi^2} \left[ \sum_i \BR_{\nu_i} \,T_{\nu_i} \,e^{2{\mu_{\nu_i}\over T_{\nu_i}}}\,
K_2\Big({m_\phi\over T_{\nu_i}}\Big) -  T_\phi \,e^{\mu_\phi\over T_\phi}\,K_2\Big({m_\phi\over T_\phi}\Big) \right]\,,
\eeq
whereas for the 3 neutrinos
\beq
\d_t n_{\nu_i} = {\Gamma_\phi m_\phi^2 \over 2 \pi^2} \,\BR_{\nu_i} \left[ 
2\, T_\phi \,e^{\mu_\phi\over T_\phi}\,K_1\Big({m_\phi\over T_\phi}\Big) 
-2\,T_{\nu_i} \,e^{2{\mu_{\nu_i}\over T_{\nu_i}}}\,K_1\Big({m_\phi\over T_{\nu_i}}\Big) \right]\,;
\eeq
\beq
\d_t \rho_{\nu_i} = {\Gamma_\phi m_\phi^3 \over 2 \pi^2} \,\BR_{\nu_i} \left[ 
T_\phi \,e^{\mu_\phi\over T_\phi}\,K_2\Big({m_\phi\over T_\phi}\Big) -
T_{\nu_i} \,e^{2{\mu_{\nu_i}\over T_{\nu_i}}}\,K_2\Big({m_\phi\over T_{\nu_i}}\Big) \right]\,,
\eeq
being $K_n(x)$ modified Bessel functions of the second kind.
The equations describing the evolution of $T$ and $\mu$ of a generic species read 
\beq
{\dd T\over \dd t} = {1\over \d_\mu n \; \d_T \rho- \d_T n\; \d_\mu \rho }
\left[ -3H\left( (p+\rho)\, \d_\mu n - n \, \d_\mu \rho \right) + \d_\mu n \, \d_t \rho -
\d_\mu \rho\, \d_t n \right]\,,
\eeq
\beq
{\dd \mu\over \dd t} = {-1\over \d_\mu n \; \d_T \rho- \d_T n\; \d_\mu \rho }
\left[ -3H\left( (p+\rho)\, \d_T n - n \, \d_T \rho \right) + \d_T n \, \d_t \rho -
\d_T \rho\, \d_t n \right]\,.
\eeq

\begin{figure}
\begin{center}
\includegraphics[scale=0.55]{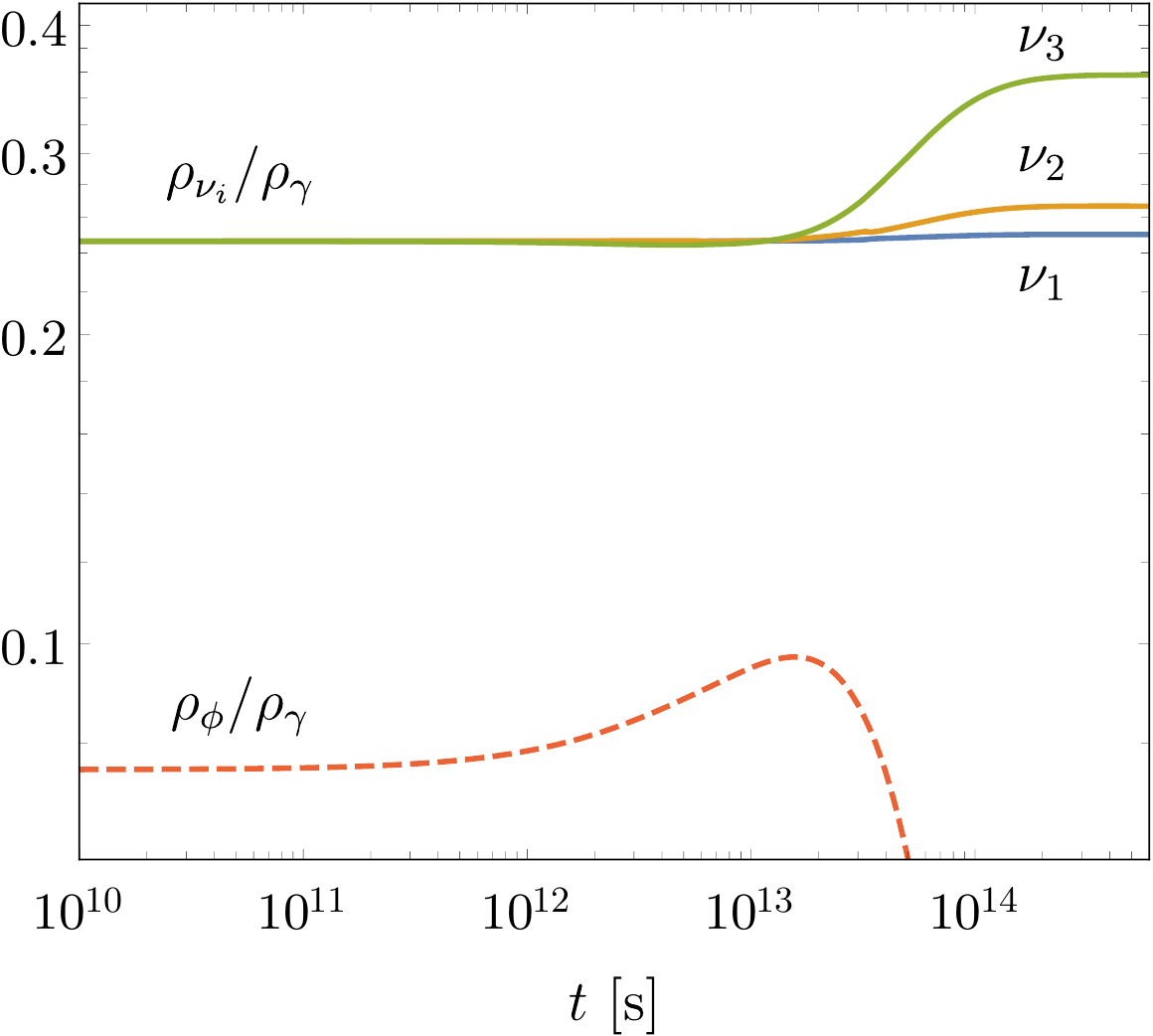} \hspace{1cm}
\includegraphics[scale=0.55]{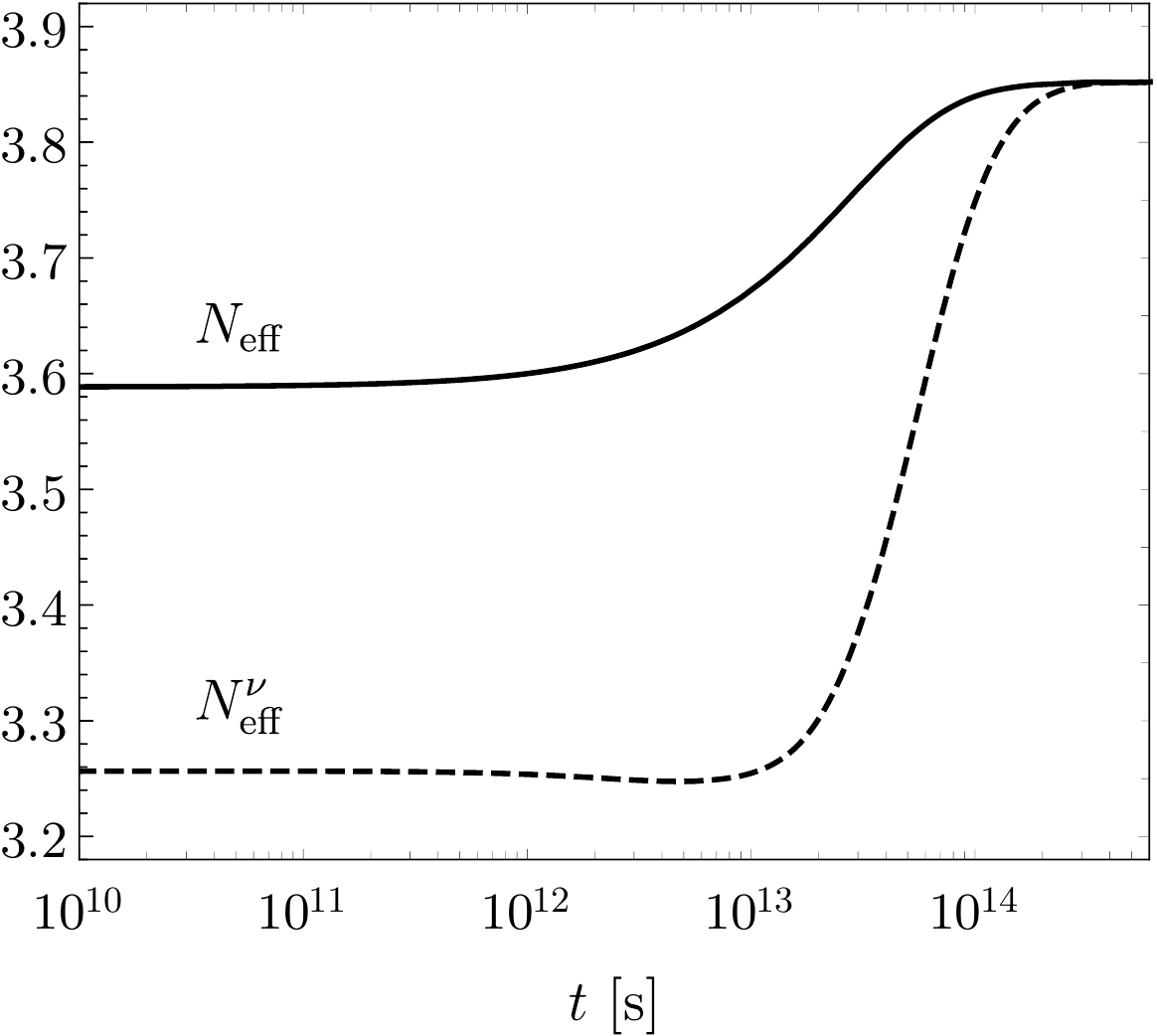} \\
\vspace{0.5cm}\includegraphics[scale=0.55]{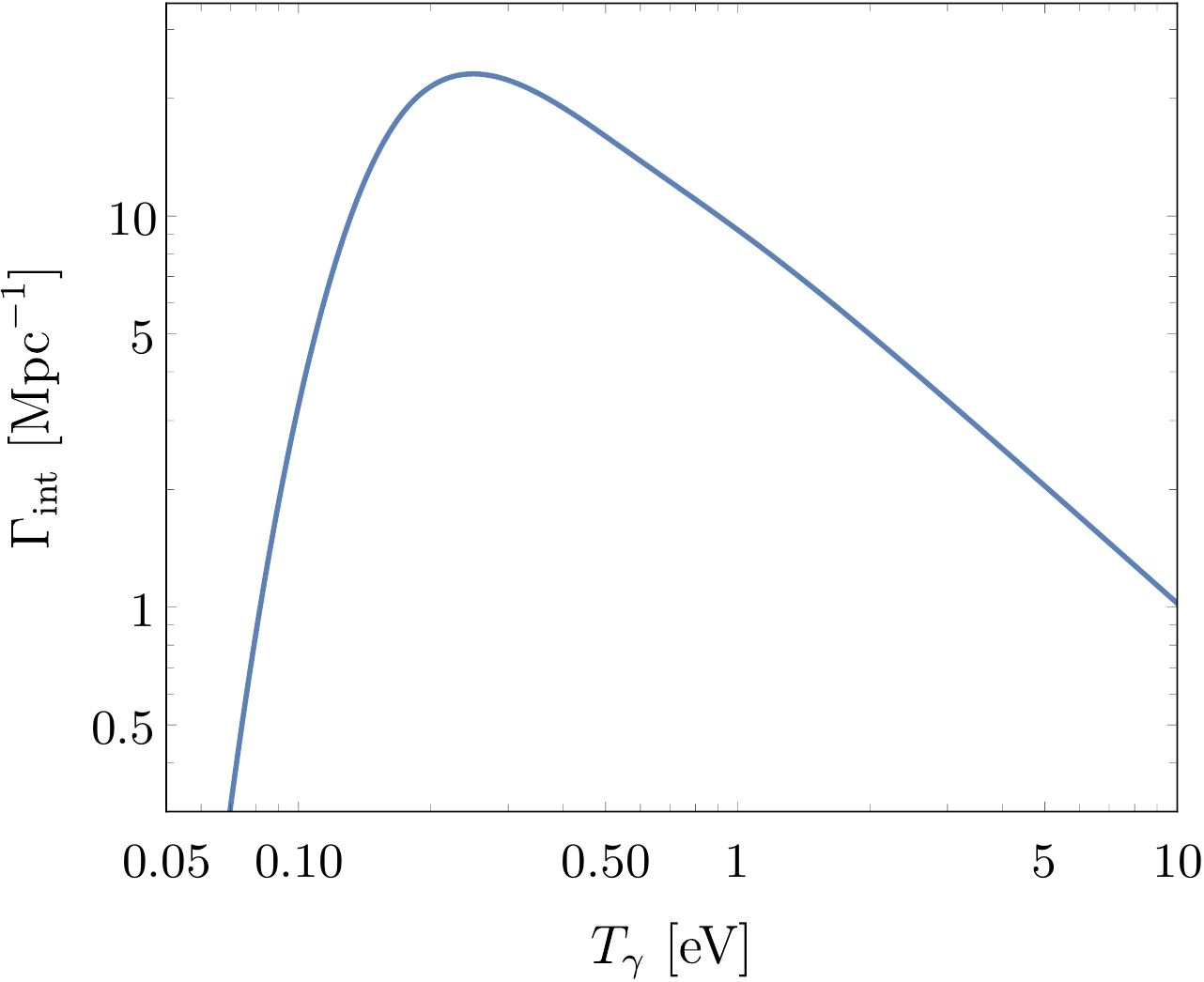}
\caption{ {\bf Left:} Ratios $\rho_{\phi}/\rho_{\gamma}$ and $\rho_{\nu_i}/\rho_{\gamma}$ for $i=1,2,3$ for $t$
between $10^{10}$ s ($T_\gamma=11$ eV) and $6\times 10^{14}$ s ($T_\gamma=0.045$ eV).
{\bf Right:} $N_{\rm eff}$ (in dashes, the contribution from neutrinos).
{\bf Bottom:} Interaction rate of neutrinos at temperatures $T_\gamma \approx m_\phi$.}
\label{fig3}
\end{center}
\end{figure}
In Fig.~{\ref{fig3} we provide the energy densities of the majoron and the three neutrino flavors relative
to $\rho_\gamma$ (left) together with $N_{\rm eff}$,
\beq
N_{\rm eff} = {8\over 7} \left( {11\over 4} \right)^{4/3} \, {\rho_\nu+\rho_\phi \over \rho_\gamma}\,.
\eeq
The plots show that  inverse decays $\,\nu_3 \bar \nu_3 \to \phi\,$
keep the energy density in the $\phi$ component of the fluid close to the equilibrium one
up to $t\approx 10 \,\tau_\phi$. Then, at temperatures below $m_\phi$ 
majorons disappear and transfer all their entropy to (mostly) $\nu_3$.
In Fig.~{\ref{fig3}-right we plot the evolution of $N_{\rm eff}$; notice that the transfer of energy between 
$\nu_{3}$ and the majorons results in a significant net increase in $N_{\rm eff}$: it goes 
from 3.58 at $T>10$ eV to 3.85 at $T<0.1$ eV. This effect, discussed in \cite{Escudero:2020dfa}, 
appears because the majoron decays into neutrinos
at temperatures below its mass, when it is mildly relativistic. In our scenario the majoron-neutrino interaction rate 
[in Fig.~\ref{fig3}-bottom, see also Eq.~(\ref{IntRate})] is 
slightly below the one required for thermal equilibrium (it corresponds to 
$\Gamma_{\rm eff}=0.8$ in the notation of \cite{Escudero:2021rfi}), which enhances the effect.

In addition, the processes $\nu_3 \bar \nu_3 \leftrightarrow \phi$ will 
damp the number of free streaming neutrinos, something that is
necessary to achieve isotropy and 
preserve the position of the peaks of the CMB acoustic oscillations \cite{Chacko:2003dt}. The 
net effect is a time-dependent modification to the growth of potential wells; in particular, these interactions
are effective
between $10\,m_\phi \ge T_\gamma \ge m_\phi/10$, implying that they will only alter CMB multipoles $\ell \le 1000$.

\begin{figure}
\begin{center}
\includegraphics[scale=0.51]{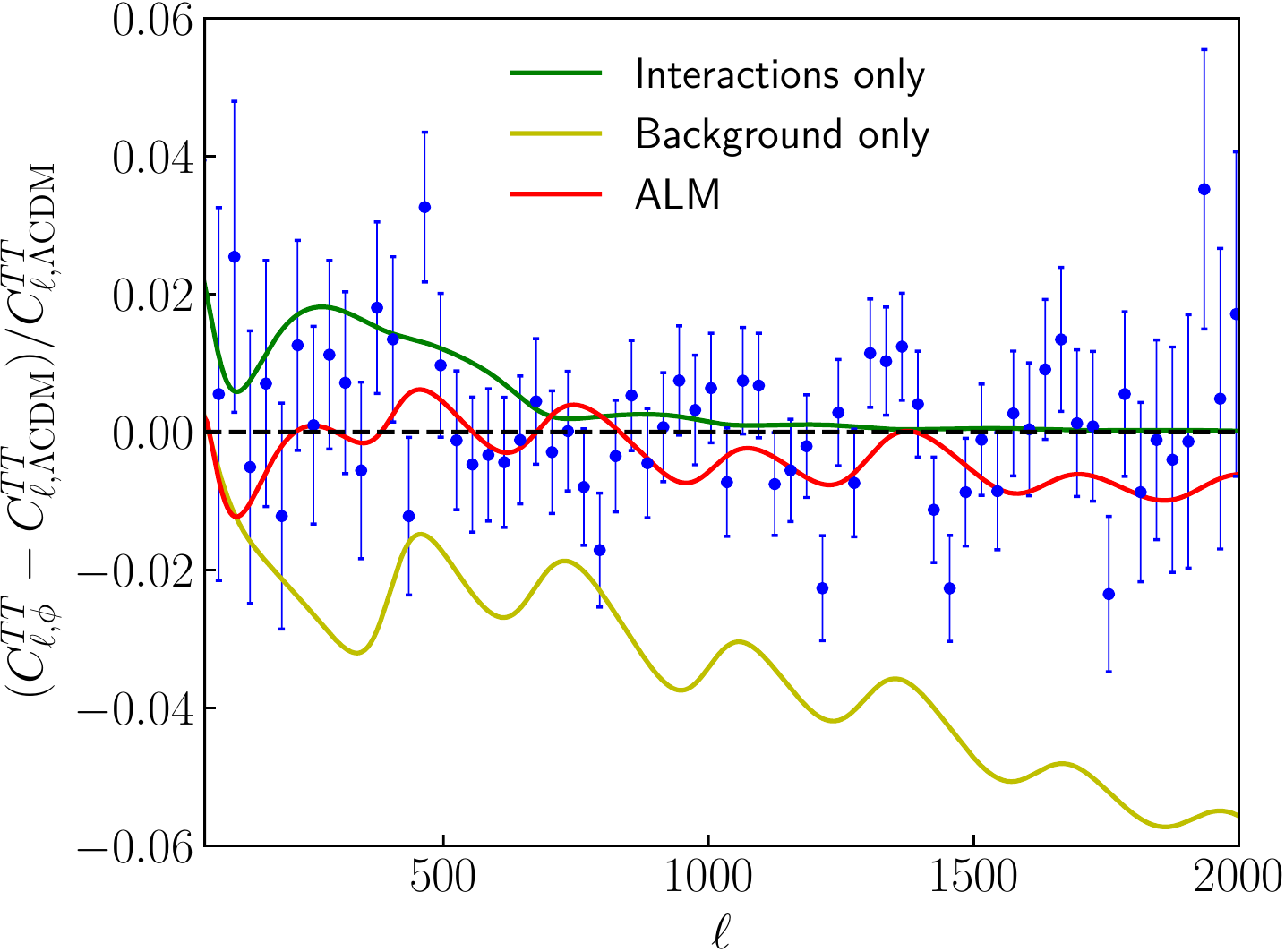}
\caption{Effect (relative to that of $\Lambda$CDM) on the TT power spectrum produced by {\it (i)} 
the $T$-dependent increase of $N_{\rm eff}$ in the ALM model; 
{\it (ii)} the damping of the $\nu$ free streaming due to interactions; and 
{\it (iii)} the total effect after redefining the cosmological parameters (see Table~\ref{table2}) in the ALM 
model.}
\label{fig4}
\end{center}
\end{figure}
Following the method and approximations\footnote{These include neglecting neutrino masses, assuming that 
neutrinos and majorons form a single coupled fluid, and using the relaxation time
approximation for their collision term \cite{Hannestad:2000gt}.}
described in \cite{Escudero:2020dfa}, we have obtained the interaction rate (in Fig.~\ref{fig3})
\beq
\Gamma_{\rm int}={1\over \rho_\nu} {\delta\rho_\nu\over \delta t}= {\Gamma_\phi \over \rho_\nu} \, \sum_i
{\rm BR}_{\nu_i}\,e^{\mu_{\nu_i}\over T_{\nu_i}} \left({m_\phi \over T_{\nu_i}} \right)^{\! 3} K_2\Big({m_\phi\over T_{\nu_i}}\Big)\,,
\label{IntRate}
\eeq
we have modified the equations for 
the time evolution of the density, velocity, shear and higher anisotropic moments of the phase space distribution in
the synchronous gauge \cite{Ma:1995ey} and we have included them in CLASS \cite{Lesgourgues:2011re,Blas:2011rf},
using MontePython \cite{Audren:2012wb,Brinckmann:2018cvx} to deduce the cosmological implications of this
scenario.

\begin{table}
\begin{center}
\footnotesize{
\begin{tabular} { |  m{2.6cm} |  m{2.6cm}|  m{2.6cm} | } 
\hline
  Parameter& $\Lambda$CDM & ALM \break $m_\phi=0.5$ eV \hspace{0.5cm}\break $\tau_\phi=3.5\times 10^{12}$ s\\ 
  \hline
$100\, \Omega_b h^2$   &  $2.242\pm 0.015$ & $2.295\pm 0.014$\\ 
$\Omega_{\rm cdm} h^2$   & $0.119\pm 0.001$ & $0.129\pm 0.001$\\ 
$100\, \theta_s$   & $1.0420\pm  0.0003$& $1.0407\pm 0.0003$\\ 
$\ln \left( 10^{10}A_s\right)$   & $3.046\pm0.015$ & $3.062\pm 0.016$ \\ 
$n_s$   & $0.967\pm 0.004$ &  $0.991\pm 0.004$ \\ 
$\tau_{\rm reio}$   & $0.055\pm 0.008$ & $0.056\pm 0.008$\\ 
   \hline
$H_0$  [km/s/Mpc] & $67.71\pm 0.44$ & $71.4\pm 0.5$ \\ 
 \hline
\end{tabular}
}
\caption{Cosmological parameters in $\Lambda$CDM and the ALM model from a combined analysis of
Planck2018 (low and high multipoles of the CMB temperature and polarization power spectra, as well as the CMB lensing likelihood, \cite{Planck:2019nip}) + BAO (BOSS, SDSS MGS \cite{BOSS:2016wmc,Ross:2014qpa,Beutler:2011hx}) data. The quoted $\Lambda$CDM values can be found in \cite{Aghanim:2018eyx}.
\label{table2}}
\end{center}
\end{table}
Our results are summarized in Fig.~\ref{fig4} and Table~\ref{table2}. 
Throughout our analysis we use the definite ALM model with $m_\phi=0.5$ eV
and $\tau_\phi=3.5\times 10^{12}$ s defined in previous sections.
We do not explore the probability density distributions of cosmological parameters for other input values, which would correspond to a different thermal history.
Fig.~\ref{fig4}  shows the effect (relative to that of $\Lambda$CDM)
on the $T$ angular power spectrum (TT) of {\it (i)} 
the $T$-dependent increase of $N_{\rm eff}$ in the ALM model; {\it (ii)} the damping of the free streaming due to the neutrino 
interactions; and {\it (iii)} the net effect 
after redefining the cosmological parameters as given in
the table. 
The first two lines are produced fixing $\Omega_b H_0^2$, $z_{\mathrm{mr}}$ and $\theta_s$ to their $\Lambda$CDM values, whereas the red line is obtained by varying the 6 cosmological parameters 
to the values given in Table~\ref{fig2}. 
The baryon to photon ratio in our model 
($100\,\Omega_b h^2=2.295\pm 0.014$ versus $2.242\pm 0.015$ in $\Lambda$CDM) may seem
inconsistent with the recent values around 
$2.20\pm 0.05$ deduced from data on the abundance of primordial Deuterium (see \cite{Pisanti:2020efz} and references therein). 
However, in the ALM model the baryon to photon ratio at the beginning of BBN (before a fraction of photons 
convert into majorons) is a $4.7\%$ smaller,
\beq
100\,\Omega_b h^2=2.186\pm 0.014\,,
\eeq
which is in perfect agreement with that data.

A comment about our choice of parameters in the ALM model is here in order. We took
$m_\phi=0.5$ eV to damp the free streaming of neutrinos and majorons 
at the {\it right} cosmological time (near recombination). 
Then we  used $g_{\phi \gamma \gamma}$ 
to convert  $6\%$ of CMB photons into ALMs, so that 
$N_{\rm eff}=3.85$ after all majorons have decayed. This value of $N_{\rm eff}$ is larger than
the ones considered in previous literature \cite{Escudero:2019gvw,Escudero:2021rfi}, and thus 
it implies an also larger expansion rate $H_0$. Finally, 
the majoron-neutrino coupling $\lambda_{\nu_3}=6.8\times 10^{-14}$ was chosen
so that $v_X\approx 900$ GeV and collider bounds on the heavy neutrinos 
are easily avoided \cite{Hernandez-Tome:2019lkb}. Notice that $\lambda_{\nu_3}$
fixes both the lifetime of the ALM and the frequency of 
neutrino-majoron interactions: its value must let the majoron decay
before recombination (a critical process to reduce the free streaming of the extra radiation during
that period) and  
imply an acceptable fit to the CMB multipoles. 

\section{Summary and discussion}
Our understanding of the neutrino sector is still work in progress. The SM admits a minimal completion of the sector 
with just the addition of the dimension-5 Weinberg operator and no new degrees of freedom at EW
energies.  Indeed, the large scale \mbox{$\Lambda\approx 10^{10}$~GeV} suppressing that operator could be explained naturally by the usual (type~I)
seesaw mechanism. However, there are other possibilities that are also well motivated and imply the presence of light exotic particles. In these
other scenarios the tiny value of the neutrino masses is justified not by a large
scale but by the breaking of a global symmetry. Ultimately, only the data may decide the right venue.

Here we have proposed one of such models and have studied its cosmological implications. 
The model is a explicit realization of the generic scenario discussed in \cite{Chacko:2003dt}.
A scalar VEV breaks an approximate global symmetry and implies
the presence of a majoron $\phi$ as the only new particle below the EW scale; a small component of the VEV also breaks a discrete $Z_3$ symmetry,
generating neutrino masses and couplings 
to the majoron. Mixing of the SM charged leptons with heavy fields may then generate a small coupling of
$\phi$ with the $\tau$ lepton and axion-like couplings of type $\phi \,\widetilde F_{\mu\nu}F^{\mu\nu}$. This is a {\it special} axion-like particle: an  
ALM that decays with an almost 100\% branching ratio into neutrinos, not into gammas. For majoron models with axion couplings to gluons see \cite{Berezhiani:1989fp,Ma:2017vdv}.

We show that  in the early universe, right after BBN, the presence of a cosmological magnetic field may drive a fraction of photons into ALMs. 
Other mechanisms to reduce $T_\gamma - T_\nu$ after BBN but before recombination include
the photon cooling by gravitational interactions with a Bose condensate of axions in \cite{Erken:2011vv} or 
by kinetic mixing with hidden photons in \cite{Jaeckel:2008fi}.

At lower temperatures,  $T_\gamma\approx m_\phi=0.5$ eV,
the ALMs enter in thermal contact with the heaviest neutrino, transfer all their entropy and define a 
scenario with $N_{\rm eff}\approx 3.85$,
with the final value of $\rho_{\nu_3}$ $50\%$ larger than that of $\rho_{\nu_1}$. This is basically the mechanism 
proposed in \cite{Escudero:2019gvw}, although we do not change significantly 
the standard value of $N_{\rm eff}$ during BBN and thus favor larger values of 
$N_{\rm eff}$ (and thus of $H_0$) at later times.

The model proposed in this work constitutes then another variation of the $\Lambda$CDM model that may
help to relax the $H_0$ tension if it persists 
(we obtain $H_0=(71.4\pm 0.5)$ km/s/Mpc) and provides an acceptable
fit for the CMB multipoles and the rest of cosmological observables. Although 
it has been argued \cite{Knox:2019rjx} that the most likely solutions to this tension are those that modify 
the expansion history of the universe just before recombination, there actually exists a large number
of plausible interpretations (see \cite{Dainotti:2021pqg} for a recent analysis of the redshift dependence
of supernova data and \cite{DiValentino:2021izs,Schoneberg:2021qvd} for a compilation of alternative solutions). 
The data on LSS from upcoming experiments will probe the $\Lambda$CDM model to new limits; 
we think that the results described here are 
an example of the effects that could be expected in scenarios that are well motivated by particle physics.

\section*{Acknowledgments}
We would like to thank Mar Bastero, Adri\'an Carmona, Mikael R. Chala, Miguel Escudero, 
Javier Olmedo, Jos\'e Santiago and Samuel Witte for discussions.
This work was partially supported by the Spanish Ministry of Science, Innovation and Universities
(PID2019-107844GB-C21/AEI/10.13039/501100011033) and by the Junta de Andaluc{\'\i}a 
(FQM 101, SOMM17/6104/UGR, P18-FR-1962, P18-FR-5057).


\end{document}